%% file: ms.tex
\documentclass{article}

\usepackage{microtype}
\usepackage{graphicx}
\usepackage{subfigure}
\usepackage{booktabs} 

\usepackage{hyperref}

\usepackage{siunitx}
\usepackage{arydshln}   
\usepackage{multirow}
\usepackage{multicol}

\newcommand{\pion}{$^{+}$}
\newcommand{\nion}{$^{-}$}

\usepackage[accepted]{icml2020}

\icmltitlerunning{Explaining Chemical Toxicity Using Missing Features}

\begin{document}

\twocolumn[
\icmltitle{Explaining Chemical Toxicity Using Missing Features}


\icmlsetsymbol{equal}{*}

\begin{icmlauthorlist}
\icmlauthor{Kar Wai Lim}{ibmsg}
\icmlauthor{Bhanushee Sharma}{rpi}
\icmlauthor{Payel Das}{ibm}
\icmlauthor{Vijil Chenthamarakshan}{ibm}
\icmlauthor{Jonathan S. Dordick}{rpi}
\end{icmlauthorlist}

\icmlaffiliation{ibmsg}{IBM Research Singapore}
\icmlaffiliation{ibm}{IBM Research AI}
\icmlaffiliation{rpi}{Rensselaer Polytechnic Institute}

\icmlcorrespondingauthor{Payel Das}{daspa@us.ibm.com}
\icmlcorrespondingauthor{Kar Wai Lim}{kar.wai.lim@ibm.com}

\icmlkeywords{Machine Learning, Contrastive Explanations, Chemical Toxicity Prediction}

\vskip 0.5in
]



\printAffiliationsAndNotice{}  

\begin{abstract}
Chemical toxicity prediction using machine learning is important in drug development to reduce repeated animal and human testing, thus saving cost and time. It is highly recommended that the predictions of computational toxicology models are mechanistically explainable. Current state of the art machine learning classifiers are based on deep neural networks, which tend to be complex and harder to interpret. In this paper, we apply a recently developed method named contrastive explanations method (CEM) to explain why a chemical or molecule is predicted to be toxic or not. In contrast to popular methods that provide explanations based on what features are present in the molecule, the CEM provides additional explanation on what features are missing from the molecule that is crucial for the prediction, known as the pertinent negative. The CEM does this by optimizing for the minimum perturbation to the model using a projected fast iterative shrinkage-thresholding algorithm (FISTA). We verified that the explanation from CEM matches known toxicophores and findings from other work.
\end{abstract}

\section{Introduction}

The dramatic increase of potential drugs in the discovery phase over the last decade has not been accompanied by a similar increase in new drug approvals; rather, approvals have decreased with a large portion of these drug candidates being rejected \citep{Hwang2016}. Poor efficacy and toxicity remain major limiting aspects of this drug discovery \citep{Hay2014}. 
Traditionally, \textit{in vivo} methods have been used to test the toxicities of these chemicals and drugs. But these are usually expensive and time consuming and hence there has been a shift towards \textit{in vitro} and to machine learning (ML) based \textit{in silico} methods. 

A variety of ML models have been applied for this task, predicting toxicities in cells (\textit{in vitro}), in animals (\textit{in vivo}), and in humans. These models take in a variety of inputs, from chemical descriptors to mechanistic biological descriptors, such as targets and biological pathways. Chemical descriptors are used more frequently, from descriptors derived from chemical and physical properties \citep{Luco1997, Abdelaziz2016}, to structural chemical fingerprints \citep{Mayr2016}, to even chemical 3D \citep{Matsuzaka2019} and 2D images \citep{Fernandez2018}. Chemical structures, even though not always directly related to the toxic effect of a molecule \citep{Alves2016}, are commonly used for screening in the drug development process \citep{Limban2018}. The type of model applied also differs, from similarity based models \citep{Ajmani2006, Chavan2015} to SVMs \citep{Cao2012}, to random forests \citep{Polishchuk2009}, and most recently towards deep learning models \citep{Mayr2016, Jimenez-Carretero2018}. This shift towards deep learning models has been pushed by their superior predictive performance~\citep{Mayr2016}, and their ability to self select significant features \citep{Tang2018}. However, though the performance of these predictive models is getting better, this shift towards more ``black-box'' models makes it increasingly unclear to explain why a molecule is predicted as toxic or not toxic. Yet this information is extremely important, not only for the drug development process in highlighting which chemical substructures to avoid while designing new molecules, but also for providing more confidence to the end-users of this information - e.g., the experimentalists, who design molecules and \textit{in vitro} toxicity measuring assays and already have expert knowledge on which chemical (sub)structures to avoid during new molecule design. Therefore, it is highly recommended for predictions of computational toxicology models to be explainable \citep{OECD_guidance}.

Many different methods have been applied to pinpoint these toxic substructures, or ``structural alerts''/``toxicophores'' in order to help explain toxicity prediction results. However, these methods have only focused on pinpointing the \textit{presence} of certain features, such as toxicophores, on explaining the resulting toxicity predictions, but they in general have not examined the affect of the \textit{absence} of these features to generate explanations. However, when defining contrastive features that are necessary for a prediction, defining both the present and absent features, that are minimal and necessary, can help provide an explanation that is more easily comprehensible for users. 
Keeping this in mind, \citet{Dhurandhar2018} have recently developed a contrastive explanations method (CEM), to identify both pertinent positive and pertinent negative features in neural networks applied to image data (MNIST handwritten digits dataset), to text data (procurement fraud dataset), and to brain activity strength data.

In our current work, we apply this model to the domain of predicting toxicity of molecules, to provide better insights into explanations of toxicity predictions. For the task of predicting toxicity from chemical structures of molecules, these contrastive features could be chemical substructures or functional groups --- for positive pertinent features these would be the minimum required substructures for a molecule's classification (structural alerts or toxicophores in the case for a toxic class), while for negative pertinent features these would be the minimum changes to the molecule that would flip the classification of a molecule, from toxic to nontoxic or vice versa. Such an explanation will expand the scope of current toxicity prediction explanations, beyond just defining toxicophores. These explanations can also provide information on how to convert a toxic molecule to a nontoxic molecule, the exact contributing chemical substructures, as well as provide information on what can make a molecule safe. 

\section{Contrastive Explanations Method (CEM)}

One major strength of the CEM is to identify missing features that contribute to an explanation, in addition to features that are present in a molecule. To illustrate, a molecule can be nontoxic because it does not have a particular toxicophore that is harmful. Such explanation is more intuitive and complete than just saying that the molecule is nontoxic because of the presence of other structures that correlate with nontoxicity.

The CEM introduces the notion of Pertinent Negative (PN) and Pertinent Positve (PP). A PN is a subset of the feature set that is necessary for a classifier to predict a given class, while a PP is the minimal subset of features whose presence gives rise to its prediction. The CEM obtains the PN and PP via an optimization problem to look for a required minimum perturbation to the model, using a projected fast iterative shrinkage-thresholding algorithm (FISTA). Refer to \citet{Dhurandhar2018} for more details. Examples on PN and PP are given in Section~\ref{SI:results}.

\section{Toxicity Prediction Model}
\label{SI:toxicity}
We apply the CEM to a deep neural network (DNN) model used for toxicity prediction. For simplicity, we focus on one specific task within the Tox21 challenge, i.e., `SR-MMP'. This particular task is chosen because it is the simplest toxicity prediction task with 95\% AUC-ROC achieved by the winner of the Tox21 challenge.\footnote{\url{https://tripod.nih.gov/tox21/challenge/leaderboard.jsp}}

The Tox21 challenge provides the results of twelve \textit{in vitro} assays that test seven different nuclear receptor signaling effects, and five stress response effects of $\sim$10,000 molecules in cells \citep{Huang2016}; these are a subset of the broader ``Toxicology in the 21st Century'' initiative that test experimentally, \textit{in vitro}, the effect of a large number of chemicals on their ability to disrupt biological pathways through high-throughput screening (HTS) techniques \citep{tox21, Tice2013, Kavlock2009}. The SR-MMP task specifically tests the ability of molecules to disrupt the mitochondrial membrane potential (MMP) in cells \citep{Huang2016}. 
To generate the features to the DNN model, we convert the SMILES representation of each molecule into a fixed length  bit vector of size 4096 using Morgan fingerprinting \cite{Rogers2010}. These bits convey information about the substructure of the molecules, for example, the 3236th bit corresponds to a substructure containing the bromine atom (see Figure~\ref{fig:pn_bits_2}). Our Morgan fingerprints encapsulate all molecular substructures up to a radius of size~2.

Since our main focus is on the explanation method, we employ a simple DNN model, one with only three hidden layers and ReLU activation function. The sizes of the hidden layers are 2048, 1024 and 512 respectively. The output layer has 2 nodes corresponding to the toxic/nontoxic labels and is activated by a softmax. 
We train the DNN model for 200 epochs with random batches of 256 molecules per training step. Binary cross entropy was chosen as the loss function and it was optimized using the ADADELTA optimizer. 
The simple DNN model achieves a test set ROC AUC of 0.8777, with true positive of 0.5754 and true negative of 0.9524, overall, the F1 score is 0.6280. Note that the results are comparable with that of \citet[][Table S14]{Liu2019}.

\section{Results}
\label{SI:results}

\paragraph{Pertinent Negative Features.}
In our toxicity prediction model, pertinent negatives (PN) are the minimum changes to the input feature of a molecule to flip its predicted label. 

\textit{Example 1: Triphenylphosphine oxide}:
For example, triphenylphosphine oxide 
was correctly predicted to be nontoxic for the SR-MMP endpoint, with a prediction probability of 0.9999. However, if we were to add some bits corresponding to the pertinent negatives, the DNN model will now predict the molecule to be toxic in SR-MMP. 

\begin{figure}[htb]
    \centering
    \hfill
    \includegraphics[width=0.44\linewidth]{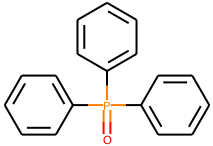}
    \hfill
    \includegraphics[width=0.32\linewidth]{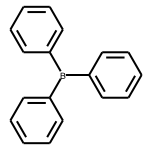}
    \hfill
    \caption{Molecular representation of triphenylphosphine oxide (left) and triphenylborane (right).}
    \label{fig:pn_mol_1}
\end{figure}%

In order to make triphenylphosphine oxide toxic, the bits to be added are 690, 876, 164, 62, 276, etc. To give an illustration of what these bits entail, we refer to other molecules in the database that have these bits and borrow the toxic parts corresponding to these bits. A selection of top 10 parts for the pertinent negatives are displayed below in Figure~\ref{fig:pn_bits_1}. Interestingly, we can see that the PN includes the trichloride (SbCl3), which is known to be very toxic \citep{Cooper2009}.

\begin{figure}[htb]
  \centering
  \includegraphics[width=0.19\linewidth]{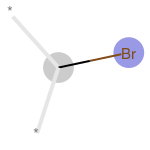}
  \includegraphics[width=0.19\linewidth]{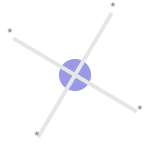}
  \includegraphics[width=0.19\linewidth]{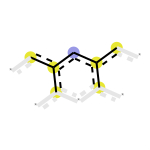}
  \includegraphics[width=0.19\linewidth]{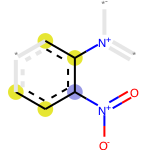}
  \includegraphics[width=0.19\linewidth]{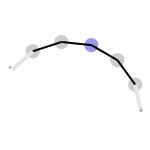}
  \includegraphics[width=0.19\linewidth]{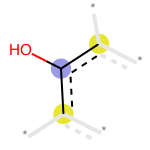}
  \includegraphics[width=0.19\linewidth]{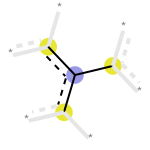}
  \includegraphics[width=0.19\linewidth]{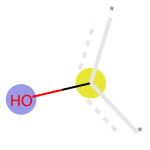}
  \includegraphics[width=0.19\linewidth]{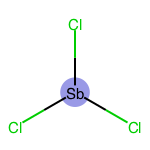}
  \includegraphics[width=0.19\linewidth]{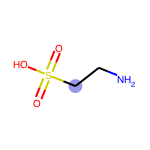}
  \caption{Pertinent Negatives (PN) for triphenylphosphine oxide, adding these parts into the molecule will make it toxic. For these substructures, the blue circle highlights the central atom, the yellow circles are part of an aromatic ring, and the gray circles are part of an aliphatic ring.}%
  \label{fig:pn_bits_1}%
\end{figure}

\textit{Example 2: Triphenylborane}:
We now look at another example where the DNN model misclassified the molecule triphenylborane to be nontoxic, where it is in fact toxic for the endpoint SR-MMP. Here, we use the pertinent negatives to identify why the model makes such false prediction.


For triphenylborane, the CEM algorithm provides only 5 pertinent negative bits, meaning that it takes less changes to flip the prediction to toxic (classification probability for toxic class changed from 0.0205 to 0.5062). Here, we can clearly see that adding of bromine atom and hydroxyl (OH) group would flip the classifier's prediction. Here, one way to explain why the classifier makes the false prediction is because such features were not observed in triphenylborane.
In fact, this molecule bears close resemblance to triphenylphosphine oxide (example 1) where the only difference is in the connecting molecule in between the aromatic rings. Thus it is not bizarre to see such a false prediction.

In contrast to existing explanation methods that provide explanation based on what is present in a molecule, the pertinent negatives instead look at what is missing from the molecule, which is an  useful and more complete explanation criteria. This is one of the main strength of using the pertinent negatives for explanation.

\begin{figure}[htb]
  \centering
  \includegraphics[width=0.19\linewidth]{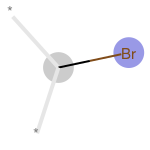}
  \includegraphics[width=0.19\linewidth]{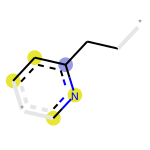}
  \includegraphics[width=0.19\linewidth]{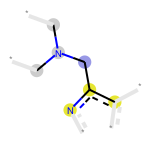}
  \includegraphics[width=0.19\linewidth]{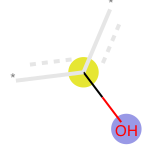}
  \includegraphics[width=0.19\linewidth]{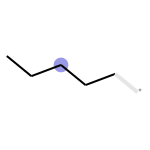}
  \caption{Pertinent Negatives (PN) for triphenylborane, adding these parts into the molecule will make it toxic.}%
  \label{fig:pn_bits_2}%
\end{figure}

\paragraph{Pertinent Positive.}

Pertinent Positives (PP) answer the question: what is the minimum required features such that a molecule 
is classified as such?

\textit{Example 3: Ethalfluralin}:
Ethalfluralin is a known herbicide for weeds. Here, the DNN model correctly predicted that the molecule is toxic on the endpoint SR-MMP. 

\begin{figure}[htb]
    \centering
    \hfill
    \includegraphics[width=0.49\linewidth]{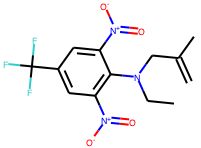}
    \hfill
    \includegraphics[width=0.39\linewidth]{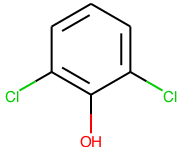}
    \hfill
    \caption{Molecular representation of ethalfluralin (left) and 2,6-dichlorophenol (right).}
    \label{fig:pp_mol_1}
\end{figure}%

\begin{figure}[htb]
  \centering
  \includegraphics[width=0.19\linewidth]{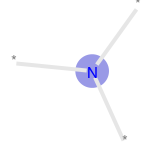}
  \includegraphics[width=0.19\linewidth]{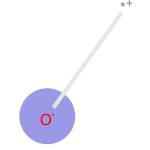}
  \includegraphics[width=0.19\linewidth]{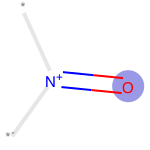}
  \includegraphics[width=0.19\linewidth]{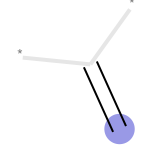}
  \includegraphics[width=0.19\linewidth]{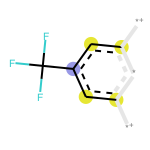}
  \caption{Pertinent Positives (PP) for ethalfluralin, which are the contributing parts for the toxic prediction of the molecule.}%
  \label{fig:pp_bits_1}%
\end{figure}

To understand why the model makes such prediction, we can inspect the pertinent positive for the molecule. In Figure~\ref{fig:pp_bits_1}, we display five pertinent positives for the prediction.

Here, we can see that one of the contributing factor to its toxicity prediction is its trifluoromethyl functional group. However, we caution that this functional group may or may not be the causal factor to its toxicity, as the CEM can only extract correlated factors rather than  causal ones.

\textit{Example 4: 2,6-Dichlorophenol}:
Next, we look at an example when the DNN model predicts a molecule incorrectly. The 2,6-dichlorophenol is corrosive and is an environmental hazard according to PubChem but it is not toxic in the SR-MMP endpoint. However, the DNN model predicts it to be toxic with a probability of 0.9999. We employ the CEM to look for its pertinent positive bits.

\begin{figure}[htb]
  \centering
  \hfill
  \includegraphics[width=0.19\linewidth]{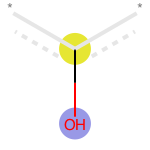}
  \hfill
  \includegraphics[width=0.19\linewidth]{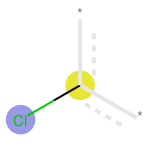}
  \hfill
  \includegraphics[width=0.19\linewidth]{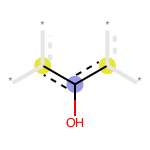}
  \hfill 
  \caption{Pertinent Positives (PP) for 2,6-dichlorophenol, which are the contributing parts for the toxic prediction of the molecule.}%
  \label{fig:pp_bits_2}%
\end{figure}

The CEM identifies three main bits for the pertinent positive, which are illustrated in the Figure~\ref{fig:pp_bits_2}. These three bits correspond to the hydroxyl (OH) functional group and the chlorine atom. The presence of these parts in the molecule resulted in the DNN model to predict the molecule to be toxic, albeit incorrectly.

\paragraph{Verifying Identified Explanations.}

\citet[][Fig 3]{Akita2018} found that the phenolic OH group is responsible for the prediction of the toxicity of the molecule tyrphostin 9 in their classification model. This is consistent with the finding of CEM, which identified the OH functional group as an explanation (in both PP and PN) for our predictions. In addition, we also found that the phenolic OH functional group solely forms the pertinent positive of tyrphostin 23 (not displayed), which is closely related to tyrphostin 9.

\paragraph{Checking with Known Toxicophores.}

Our model's explanations, through identifying PP and PN substructures, can pinpoint structural alerts --- substructures in molecule that are required for a toxic label or substructures that can flip the predicted label to a toxic label. 

The SR-MMP task measures the ability of molecules to disrupt the mitochondrial membrane potential in cells.  Weakly acid groups are known to uncouple this membrane potential, with the phenolic OH being the most common weakly acid group \citep{Terada1990}. Our model identifies phenolic OH groups as PP or PNs to for multiple molecules, consistent with this work (Table~\ref{tbl:tox_SA}). 

Further, through past experiments and testing, various substructures have already been known to act as structural alerts / toxicophores for different toxic endpoints. A commonly used experiment is the Ames test that tests for the mutagenicity of chemicals \textit{in vitro} in cells, general structural alerts from these have already been curated in past studies and is a common benchmark in identifying toxicophores in molecules \citep{Kazius2005, Kazius2006, Yang2017}. Our model was able to identify several of these general toxicophores that are connected to mutagenicty, e.g., aliphatic halides, polycylic aromatic systems, aromatic nitros \citep{Kazius2006}, as both PP and PN to either explain the toxicity of a molecule or to turn a molecule into a toxic one\,(Table\,\ref{tbl:tox_SA}). 

\input{tbl_SA}

\section{Conclusion}

Applying the recently proposed CEM \citep{Dhurandhar2018} to the task of predicting toxicity of molecules, in particular the SR-MMP task, we were able to demonstrate the ability to explain toxic predictions both from structural alerts as well as examining the substructures that are missing that can switch a molecule to be toxic. This is helpful in terms of both identifying toxicophores that might need to be avoided in designing a molecule, as well as giving an indication of how to convert a molecule to be toxic, and thus reversely nontoxic. The toxicophores extracted in such a manner, were consistent with both known experimental toxicophores and past models of identifying substructures. 

To note, even though a molecule can be classified as toxic or nontoxic by the SR-MMP endpoint, this is only measuring its toxicity in regards to this one endpoint - whether the molecule can disrupt the mitochondrial membrane potential. Thus this cannot necessarily predict if the molecule will be toxic for other endpoints, but can still give an indication of this. This can be the reason for seemingly contradicting known toxicities and toxicity for the SR-MMP endpoint, e.g., 2,6-dichlorophenol is corrosive and an environmental hazard, but not toxic in the SR-MMP endpoint. Thus, it is important to be cognisant of the definition of toxicity we are examining for our explanations.  


Our PNs are general examples of the minimum required substructures that can flip a molecule's label to toxic, thus \textit{missing} substructures that cause the predicted label to be nontoxic. This is a good starting point to associate a particular substructure to toxicity, however for a more complete explanation, the next step would be to pinpoint the location to which these substructures can be added to the molecule to create chemically meaningful molecules.

Overall, the CEM is able to help discern both the missing and contributing structural alerts for toxicity that are consistent to known toxicophores. Further work will extend to identification  of causal factors to molecular toxicity. 




\bibliography{biblio}
\bibliographystyle{icml2020}





\end{document}

%% file: tbl_SA.tex
\begin{table}[t]
    \begin{center}
    \vspace{-0.1in}
    \caption{
        Structural alerts identified by the CEM match known general toxicophores (see \citealp{Kazius2005}).
    }%
    \label{tbl:tox_SA}
    \scalebox{0.85}{
        \begin{tabular}{ccccc}
            \toprule
            {General}
            & \multicolumn{2}{c}{~~~~~~Pertinent Negative}
            & \multicolumn{2}{c}{Pertinent Positive}
            \\
            {Toxicophore} & {SA} & {Eg} & {SA} & {Eg} 
            \\
            \midrule 
            \multirow{2}{*}{Phenolic OH} & 
            {\footnotesize -cc(O)c-} & (1) & 
            \multirow{2}{*}{\footnotesize -cO} & \multirow{2}{*}{(3)}
            \\
            & {\footnotesize -cO} & (1, 2) & &
            \\
            \hline
            \multirow{1}{*}{Aliphatic halides} & 
            {\footnotesize Br} & (1, 2) & 
            {\scriptsize -cc(c(F)(F)F)cc-} & (3)
            \\
            (Cl, Br, I, F) & {\footnotesize Br} & (2) & 
            {\footnotesize -c(Cl)} & (4)
            \\
            \hline
            \multirow{1}{*}{Polycyclic} & 
            \multirow{2}{*}{\footnotesize -c(ccc(c)c)c-} \hspace{-2mm} & \multirow{2}{*}{(1)} & & 
            \\
            aromatic system 
            \\
            \hline
            \multirow{1}{*}{Aromatic nitro} & 
            {\scriptsize -ccc(N\pion(O\nion)=O)c(N\pion)c-} \hspace{-4mm} & (1) & 
            {\scriptsize N, O\nion, N\pion=O} \hspace{-2mm} & (3) 
            \\
            \bottomrule
        \end{tabular}
    }
    \end{center}
    \vspace{-0.25in}
\end{table}